# Enhancement of Perpendicular Magnetic Anisotropy and Transmission of Spin-Hall-Effect-Induced Spin Currents by a Hf Spacer Layer in W/Hf/CoFeB/MgO Layer Structures

Chi-Feng Pai,[1] Minh-Hai Nguyen,[1] Carina Belvin,[2] Luis Henrique Vilela-Leão,[1] D. C. Ralph,[1,3] and R. A. Buhrman[1*]

[1]Cornell University, Ithaca, New York, 14853, USA

[2]Department of Physics, Wellesley College, Massachusetts, 02481, USA

[3]Kavli Institute at Cornell, Ithaca, New York, 14853, USA

We report that strong perpendicular magnetic anisotropy of the ferromagnetic layer in a W/CoFeB/MgO multilayer structure can be established by inserting a Hf layer as thin as 0.25 nm between the W and CoFeB layers. The Hf spacer also allows transmission of spin currents generated by an in-plane charge current in the W layer to apply strong spin torque on the CoFeB, thereby enabling current-driven magnetic switching. The antidamping-like and field-like components of the spin torque exerted on a 1 nm CoFeB layer are of comparable magnitudes in this geometry. Both components originate from the spin Hall effect in the underlying W layer.

---

[*] E-mail: rab8@cornell.edu

The ability of spin-polarized electron currents and pure spin currents to manipulate magnetism at the nanoscale via spin transfer torque (STT) has opened a broad range of opportunities in metallic spintronics. Recently the discovery of a strong perpendicular magnetic anisotropy (PMA) at CoFeB/MgO interfaces has enabled a substantial reduction in the bias current required for STT switching of the free layer of MgO magnetic tunnel junctions, advancing STT magnetic random access memory towards commercialization.[1] Another significant development has been the determination of large spin Hall effects (SHE)[2-4] in certain high-atomic-number normal metal (NM) films. This has enabled new three-terminal device approaches wherein an in-plane charge current flowing in a thin film microstrip can exert an antidamping-like torque on an adjacent ferromagnetic (FM) layer that is strong enough to realize out-of-plane magnetic switching,[5,6] in-plane magnetic switching,[7,8] magnetic oscillations,[9,10] domain wall depinning,[11] and control of chiral domain wall displacement.[12,13] Further progress will be made if the strength of the PMA in spintronics structures that employ CoFeB/MgO MTJs can be substantially increased, and if CoFeB layers with enhanced PMA can be successfully combined with more of the full range of materials that exhibit strong SHEs. Recent progress towards those objectives include the demonstrations that Hf under-layers can materially enhance the PMA of thin (~ 1 nm) CoFeB layers that are covered by MgO,[14] and also that the controlled nitrogen doping of a Ta under-layer can enhance the PMA of CoFeB/MgO bilayers.[15]

Here we report an investigation that achieves a strong PMA by utilizing thin (0.25 to 5 nm) layers of Hf inserted between a CoFeB/MgO bilayer and W, which has a particularly large spin Hall angle $\left|\theta_{SH}^{\beta\text{-W}}\right| \approx 0.3$.[8] While Hf/CoFeB/MgO multilayers without a W layer produce negligible current-induced spin-orbit torques, W/Hf/CoFeB/MgO structures exhibit strong torques, with comparable magnitudes for the antidamping-like and field-like components. These results clarify that the origin of both components of current-induced torque in these structures is from the W layer, and they demonstrate a strategy for using different materials to generate the PMA and spin-orbit torques in multilayer structures so that the PMA and spin-orbit torques can be optimized separately.

The multilayers discussed in this paper were prepared by sputter deposition from 2-inch planar magnetron sources onto thermally-oxidized Si substrates. The base pressure of the sputtering system was less than $4 \times 10^{-8}$ Torr. The DC sputtering conditions for the metal layers were: Ar pressure of 2 mT and DC power of 30 watts with a deposition rate of ~ 0.01 nm/s, while the RF sputtering conditions for the MgO were: Ar pressure of 2 mT and RF power of 100 watts with a deposition rate of ~ 0.004 nm/s. A Ta capping layer (1 nm) was employed to protect the MgO layer from degradation due to exposure to the atmosphere. We prepared both W/Hf/Co$_{20}$Fe$_{60}$B$_{20}$/MgO/Ta multilayers, the main focus of this investigation, and also Hf/Co$_{20}$Fe$_{60}$B$_{20}$/MgO/Ta multilayers for control experiments. The multilayers were patterned into Hall bars with dimensions $5 \times 20$ $\mu$m$^2$ for anomalous Hall voltage measurements. Unless otherwise noted, all samples presented in this paper went through post-fabrication annealing at 300°C for 30 minutes in vacuum to enhance the PMA. The resistivities of our W, Hf, and CoFeB layes were determined by resistance versus thin film thickness measurements, giving $\rho_W = 200$ $\mu\Omega$cm, $\rho_{Hf} = 80$ $\mu\Omega$cm, and $\rho_{CoFeB} = 150$ $\mu\Omega$cm.

In Fig. 1 we show results of the measurement of: (a) the magnetic moment, and (b) the effective magnetic anisotropy energy $K_{eff}$ of the two different types of multilayers using vibrating sample magnetometry (VSM), with the CoFeB thickness $t_{CoFeB}$ ranging from 0.4 to 2 nm. The CoFeB thickness dependence of the magnetic moment (per area) indicates that $M_s \approx 1240$ emu/cm$^3$ for all the films, and there is no discernible "magnetic dead layer," for either case. This is consistent with previous work with Hf thin film under-layers[14] but is quite distinct from reports of dead layers of up to 0.5 nm in extent for Ta/CoFeB/MgO multilayers,[1,16] although this latter result does not appear to be universal for Ta/CoFeB,[14] which suggests that the presence or absence of a magnetic dead layer may be process, as well as material, dependent.

The standard expression for the thickness dependence of the effective anisotropy of a film with both an interfacial anisotropy energy density $K_s$ and a bulk anisotropy energy density $K_v$ is[17]

$$K_{eff} t_{CoFeB} = \left(K_v - 2\pi M_s^2\right) t_{CoFeB} + K_s . \tag{1}$$

The results shown in Fig. 1(b) indicate that for the W(4)/CoFeB($t_{CoFeB}$)/MgO(1.6)/Ta(1)

[thickness in nanometers] system $K_{eff}t_{CoFeB}$ is linear as a function of $t_{CoFeB}$, with the CoFeB in these multilayers exhibiting in-plane anisotropy, $K_{eff} < 0$, over the entire CoFeB thickness range studied. The interfacial anisotropy indicated by the linear fit of the data to Eq. (1) is small, $K_s = 0.02$ erg/cm$^2$. However, there is a strong positive (out-of-plane), and constant, volume anisotropy energy density $K_v = 7.5 \times 10^6$ erg/cm$^3$, similar to that recently reported for annealed Co$_{20}$Fe$_{60}$B$_{20}$ layers sputter deposited on Ta,[15] but very different from the report of $K_v \approx 0$ from a different study of Co$_{40}$Fe$_{40}$B$_{20}$ layers formed on a Ta underlayer.[14] We also note that in the former work the use of an N-doped Ta under-layer reduced $K_v$ by 60% or more while enhancing $K_s$.[15] Since the magnetostriction constant of Co$_{20}$Fe$_{60}$B$_{20}$ is reported to be positive[18] and since it is established that W films sputter-deposited at low pressure, ≤ 1 Pa, are usually in a state of compressive stress,[19] we tentatively conclude that this positive volume anisotropy is the result of compressive stress from the W/CoFeB interface overcoming the tensile stress contribution expected from a crystalline ordered CoFeB/MgO interface.

The insertion of 1 nm of Hf between the W and the CoFeB has a dramatic effect, in that the CoFeB then (see Fig. 1(b)) exhibits PMA ($K_{eff} > 0$) for $t_{CoFeB}$ ranging from 0.4 nm to 1.7 nm. The fit to the linear range (1.7 nm to 2 nm) of the data indicates $K_s = 1.9$ erg/cm$^2$ together with a small *negative* volume anisotropy term $K_v = -1.1 \times 10^6$ erg/cm$^3$. This difference in anisotropy is likely attributable in part to the high negative formation enthalpy of Hf borides particularly in comparison to that of W borides,[20,21] with the former making Hf an effective B "sink," enabling better crystallization of the CoFe/MgO interface and the enhancement of $K_s$. Below 1.7 nm the variation of $K_{eff}t_{CoFeB}$ becomes sub-linear with $t_{CoFeB}$, attaining a maximum of $K_{eff}t_{CoFeB} \approx 0.4$ erg/cm$^2$ for $t_{CoFeB} = 1.1$ nm, and then decreasing towards 0 as $t_{CoFeB} \to 0.3$ nm. While a similar non-linear behavior of $K_{eff}t_{CoFeB}$ in the thin CoFeB regime for Ta/CoFeB/MgO multilayers has been attributed to intermixing of the two metals,[16] this does not appear as plausible for our multilayers given the absence of a significant magnetic dead layer. Instead we suggest that thickness dependent stress effects originating from the base layer are the origin of the reduction of $K_s$ as $t_{CoFeB}$ decreases.

To determine the strength of the spin-orbit effective fields that are generated by an in-plane current in the W/Hf/CoFeB/MgO multilayers, we employed patterned Hall bar structures (Fig. 2(a)). We performed non-resonant magnetization-tilting measurements[22-24] by applying a small amplitude low frequency (100 Hz) alternating current (AC) ($I_{AC} = 1.1$ mA) through the device and simultaneously sweeping a static in-plane magnetic field parallel to or perpendicular to the current direction ($H_L$ and $H_T$ in Fig. 2(a)). The AC-induced effective field amplitude $H_{L(T)}$ along a given in-plane field direction can be determined as[23]

$$H_{L(T)} = -2 \frac{\partial V_{2\omega}/\partial H_{L(T)}}{\partial^2 V_{\omega}/\partial H_{L(T)}^2}. \qquad (2)$$

$V_\omega$ and $V_{2\omega}$ represent first (in-phase) and second harmonic (90° out-of-phase) anomalous Hall effect (AHE) voltage signals, which were recorded by lock-in techniques. Due to the very strong PMA in our multilayer structures, which typically exhibited out-of-plane coercive fields $H_c >$ 500 Oe (see Fig. 2(b)), we could neglect a small signal generated by the planar Hall effect (PHE), unlike the case for devices with weaker PMA where the PHE contribution must be taken into account.[24,25]

Fig. 2 (c-f) shows the field dependence of first and second harmonic AHE voltage signals along both the longitudinal and transverse directions for a Hall bar device with a W(4)/Hf(1)/CoFeB(1)/MgO(1.6)/Ta(1) multilayer. By fitting the first harmonic signals to $V_\omega = \Delta V_{AHE}\left[1 - 0.5\left(H_{L(T)}/H_{an}\right)^2\right]$,[23] we estimate that the anisotropy field of this device is $H_{an} \approx 8200$ Oe ($\Delta V_{AHE}$ is the anomalous Hall voltage difference between the $+M_z$ and $-M_z$ states), in good accord with a VSM measurement on an extended film that yielded $H_{an} \approx 8000$ Oe. The linear dependence of $V_{2\omega}$ with respect to the external field as shown in Fig. 3(c) and (d) is attributable to the current-induced effective fields, as previously studied in Pt and Ta based PMA systems.[22,23] Using Eq. (2), we determine $H_L/J_e^W = (6.6 \pm 0.2) \times 10^{-6}$ Oe/(A/cm$^2$) and $H_T/J_e^W = -(8.4 \pm 0.2) \times 10^{-6}$ Oe/(A/cm$^2$) where $J_e^W$ is the estimated amplitude of the AC current density in the W layer. In sharp contrast, the same set of measurements applied to a Hf(5)/CoFeB(1)/MgO(1.6)/Ta(1) Hall-bar device, for which $H_{an} \approx 9100$ Oe and $H_c \approx 960$ Oe, yield effective fields of negligible strength:

$H_L / J_e^{Hf} = (1.6 \pm 0.7) \times 10^{-7}$ Oe/(A/cm$^2$) (~40 times smaller than W/Hf case) and

$H_T / J_e^{Hf} = -(3.7 \pm 0.4) \times 10^{-7}$ Oe/(A/cm$^2$), where $J_e^{Hf}$ is the current density in the Hf layer. In this sample without a W layer the current-induced transverse field is approximately the same amplitude, and the same sign, as the magnetic field expected simply from Ampere's Law.

These results are a clear demonstration that a strong PMA does not necessarily correlate with the existence of strong current-induced spin-orbit torques (or effective fields) originating from the interfaces responsible for the PMA. Instead the straightforward inference is that for the W/Hf based multilayers the effective fields arise from the transverse spin current $J_s = (\hbar/2e)\theta_{SH} J_e^W$ generated by the spin Hall effect in the underlying W layer. To confirm this we used the same method to determine the effective fields in W(4)/Hf($t_{Hf}$)/CoFeB(1)/MgO(1.6)/Ta(1) samples with $t_{Hf}$ = 0.25, 0.5, 1, 2, and 4 nm. (It was necessary to anneal the $t_{Hf}$ = 0.25 nm sample at 350°C for 1 hour to obtain PMA comparable to other samples.) In Fig. 3(a) we plot the resulting values of $H_L / J_e^W$ and $H_T / J_e^W$ as a function of $t_{Hf}$. We attribute the decrease of $H_L$ with $t_{Hf}$ to the attenuation of the spin current from the W layer due to spin relaxation in the Hf spacer.

If $H_L$ corresponds to a (anti)damping-like (Slonczewski-like) torque $\vec{\tau}_d = \vec{m} \times \vec{H}_L$ exerted on the FM moment $\vec{m}$ by the spin current generated in the W then we can define an *effective* spin Hall angle of the W/Hf bilayer, $\theta_{SH}^{eff}$, taking the spin current attenuation by the Hf layer into account, such that[26]

$$H_L / J_e^W = \hbar \theta_{SH}^{eff} / 2eM_s t_{CoFeB}. \tag{3}$$

Fig. 3(b) shows the $t_{Hf}$ dependence of $|\theta_{SH}^{eff}|$ as determined by using Eq. (3). If we make the simplifying assumption[27] that there is no spin accumulation at the Hf/CoFeB interface then the spin current density $J_s$ entering the ferromagnetic layer can be expressed as

$$J_s(t_{Hf}) = \frac{J_{SHE}^W}{\cosh(t_{Hf}/\lambda_{s,Hf}) + (\rho_{Hf}\lambda_{s,Hf}/\rho_W\lambda_{s,W})\sinh(t_{Hf}/\lambda_{s,Hf})}, \tag{4}$$

where $J_{SHE}^W$, $\lambda_{s,Hf}$ and $\lambda_{s,W}$ represent the spin current from the W, the spin diffusion length of Hf, and the spin diffusion length of W respectively. In the limit of $\rho_{Hf}\lambda_{s,Hf} \ll \rho_W\lambda_{s,W}$, this results in

$$\theta_{SH}^{eff}(t_{Hf}) = \theta_{SH}^{eff}(0) \, \text{sech}(t_{Hf}/\lambda_{s,Hf}). \tag{5}$$

A fit of Eq. (5) to the data (see Fig. 3b) yields $\lambda_{s,Hf} \approx 1.5\,\text{nm}$ and $|\theta_{SH}^{eff}| = 0.34 \pm 0.05$ at $t_{Hf} = 0$, which is quite close to the previously reported result $|\theta_{SH}^{\beta-W}| = 0.33 \pm 0.06$ for a 5.2 nm W layer obtained from spin-torque ferromagnetic resonance and in-plane magnetization switching measurements.[8]

The decrease of $H_T / J_e^W$ with increasing $t_{Hf}$ shown in Fig. 3(a) is qualitatively similar to that of $H_L / J_e^W$ for $t_{Hf} > 1$ nm and thus the origin of this transverse effective field also appears clearly to be the SHE in the W. Such a *nonlocal* effective field has also been observed recently in Py/Cu/Pt devices having in-plane magnetic anisotropy.[28] The SHE has been shown[29,30] to be capable of generating a field-like torque on an adjacent FM (manifesting as an effective $H_T$) as well as an antidamping-like torque (an effective $H_L$), with the amplitudes of the two torques correlated to the imaginary and the real parts of interfacial spin mixing conductance $G^{\uparrow\downarrow}$, respectively. Therefore the similar magnitudes of $H_L$ and $H_T$ for $t_{Hf} > 1$ nm suggest that $\text{Re}(G^{\uparrow\downarrow})$ and $\text{Im}(G^{\uparrow\downarrow})$ are comparable for the Hf/CoFeB(1 nm) interface. When $t_{Hf} \leq \lambda_{s,Hf}$, on the other hand, $H_T$ unlike $H_L$ *decreases* rapidly with decreasing $t_{Hf}$. We interpret this as an indication that $\left|\text{Im}(G^{\uparrow\downarrow})\right|^{W|CoFeB} < \left|\text{Im}(G^{\uparrow\downarrow})\right|^{Hf|CoFeB}$, because the spin-torque properties of the W/Hf($t_{Hf}$)/CoFeB multilayer should cross over smoothly to those of W/CoFeB as $t_{Hf}$ decreases below $\lambda_{s,Hf}$ to 0. This materials difference is reasonable because a nonzero value of $\left|\text{Im}(G^{\uparrow\downarrow})\right|$ results from spins that precess upon reflection from the CoFeB to have a component in the longitudinal direction and then relax in the Hf or W. Because of the higher resistivity (increased scattering) in W compared to Hf, in a W/CoFeB sample reflected spins are more likely to be scattered back into the CoFeB and undergo relaxation there rather than in the normal metal, with the result being a decrease in both $\left|\text{Im}(G^{\uparrow\downarrow})\right|$ and $H_T$.

Finally to confirm the strength of the current-induced antidamping torque (the component associated with $H_L$) even in the presence of the Hf spacer layer, we performed current-induced switching studies of W(4)/Hf(1)/CoFeB(1)/MgO(1.6)/Ta(1) Hall-bar devices. In Fig. 3(c) and (d), we show that in the presence of an in-plane external field $H_x$ along the in-plane direct current (DC) direction, the device can be switched between $+M_z$ and $-M_z$ with $|I_{DC}| \approx 3$ mA. Upon reversal of the applied in-plane field direction, the polarity of the current-induced switching loop also reverses, fully consistent with SHE induced spin-torque switching.[6] We do find that to obtain reliable switching $H_x \geq 50$ Oe is required. Recent work[31] has shown that such a $H_x \geq H_{x,\min}$ requirement arises from the presence of a Dzyaloshinskii-Moriya interaction (DMI)[12,13] at the FM-NM interface. This imposes a fixed chirality on the walls of a nucleated sub-volume domain that the applied field has to modify so that the spin Hall torque can assist the thermally activated depinning of the domain walls, and thus drive domain expansion and complete reversal. The lower $H_{x,\min}$ here in comparison to that for the former work, 50 Oe vs. 100 Oe, indicates that the DMI at the CoFeB/Hf interface, while significant, is about 50% weaker than for Pt/Co.

In summary, the measurements reported here of current-induced torque and magnetic switching for W/Hf composite layer devices, which combine the strong SHE of W with the strong PMA and an DMI caused by the Hf insertion layer, extend the potential of spin-orbit torque studies and applications beyond simple Pt- and Ta-based PMA systems in which the same heavy-metal material must enable both the SHE and the PMA. Perhaps equally important, our observation of strong current-induced torques, with large values for both the antidamping-like and field-like components, for samples in which a spacer layer with no significant $|\theta_{SH}|$ is used to physically separate a NM with a large $|\theta_{SH}|$ from the FM layer, allows us to conclude that both components of spin-torque in this system originate in the W layer, and not in the magnetic layer as might be expected in Rashba-based models.[5,32] These and further studies should lead to better understanding of how to engineer more efficient spin-orbit torque devices.

**Acknowledgements**

The authors thank Junbo Park, Yongxi Ou and Praveen Gowtham for fruitful discussions. This work was supported in part by NSF/MRSEC (DMR-1120296) through the Cornell Center for Materials Research (CCMR), ONR, and ARO. We also acknowledge support from the NSF through use of the Cornell Nanofabrication Facility/NNIN, and support under the National Council for Scientific and Technological Development (CNPq) - Brazil for L. H. V-L.

**Figure Captions**

FIG. 1. (Color online) CoFeB thickness ($t_{CoFeB}$) dependence of (a) magnetic moment per unit area and (b) effective magnetic anisotropy energy (in terms of $K_{eff} t_{CoFeB}$). Red circles represent results from W(4)/CoFeB($t_{CoFeB}$)/MgO(1.6)/Ta(1) samples and blue squares represent results from W(4)/Hf(1)/CoFeB($t_{CoFeB}$)/MgO(1.6)/Ta(1) samples.

FIG. 2. (Color online) (a) Schematic illustration of the W/Hf/CoFeB/MgO hall-bar device. (b) Hall resistance hysteresis loop taken from the W(4)/Hf(1)/CoFeB(1)/MgO(1.6)/Ta(1) sample with an out-of-plane external field $H_\perp$. (c, d) First and (e, f) second harmonic signals with sweeping external field (c, e) parallel to in-plane current direction and (d, f) perpendicular to in-plane current direction. Solid squares and open triangles represent data taken from the sample in magnetization-up state ($+M_z$) and magnetization-down state ($-M_z$), respectively.

FIG. 3. (Color online) (a, b) Hf insertion layer thickness dependence of the (a) longitudinal and transverse effective fields (per current density in the W layer) and the (b) effective spin Hall angle from W(4)/Hf($t_{Hf}$)/CoFeB(1)/MgO(1.6)/Ta(1) samples. (c, d) In-plane current induced magnetization switching in W(4)/Hf(1)/CoFeB(1)/MgO(1.6)/Ta(1) sample with a fixed in-plane

longitudinal magnetic field $H_\mathrm{L}$ (c) parallel or (d) anti-parallel to the in-plane current direction.

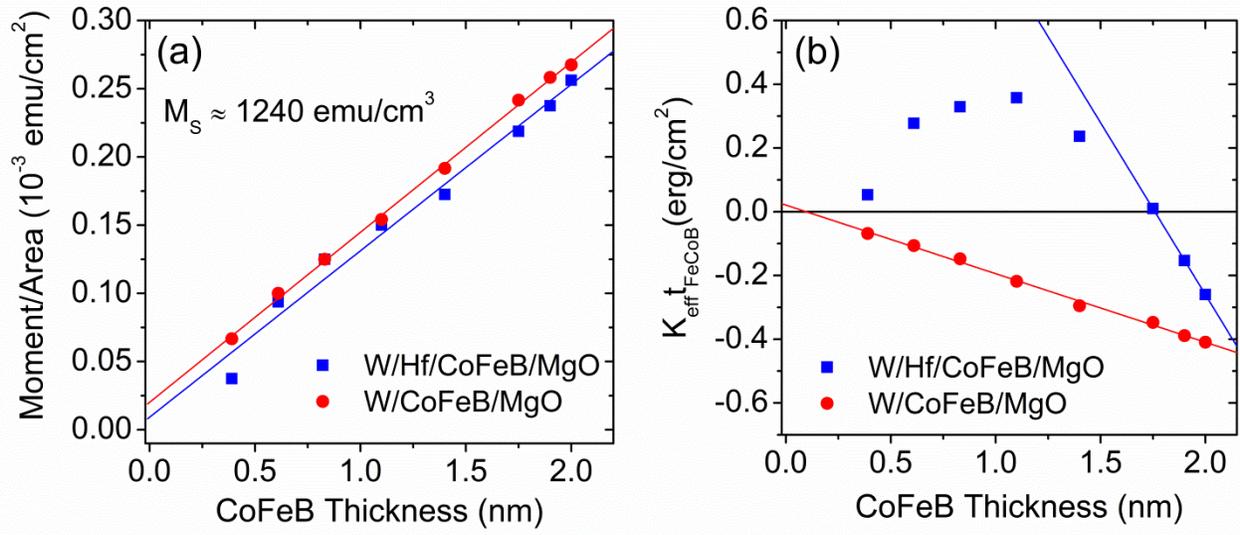

**FIG. 1**

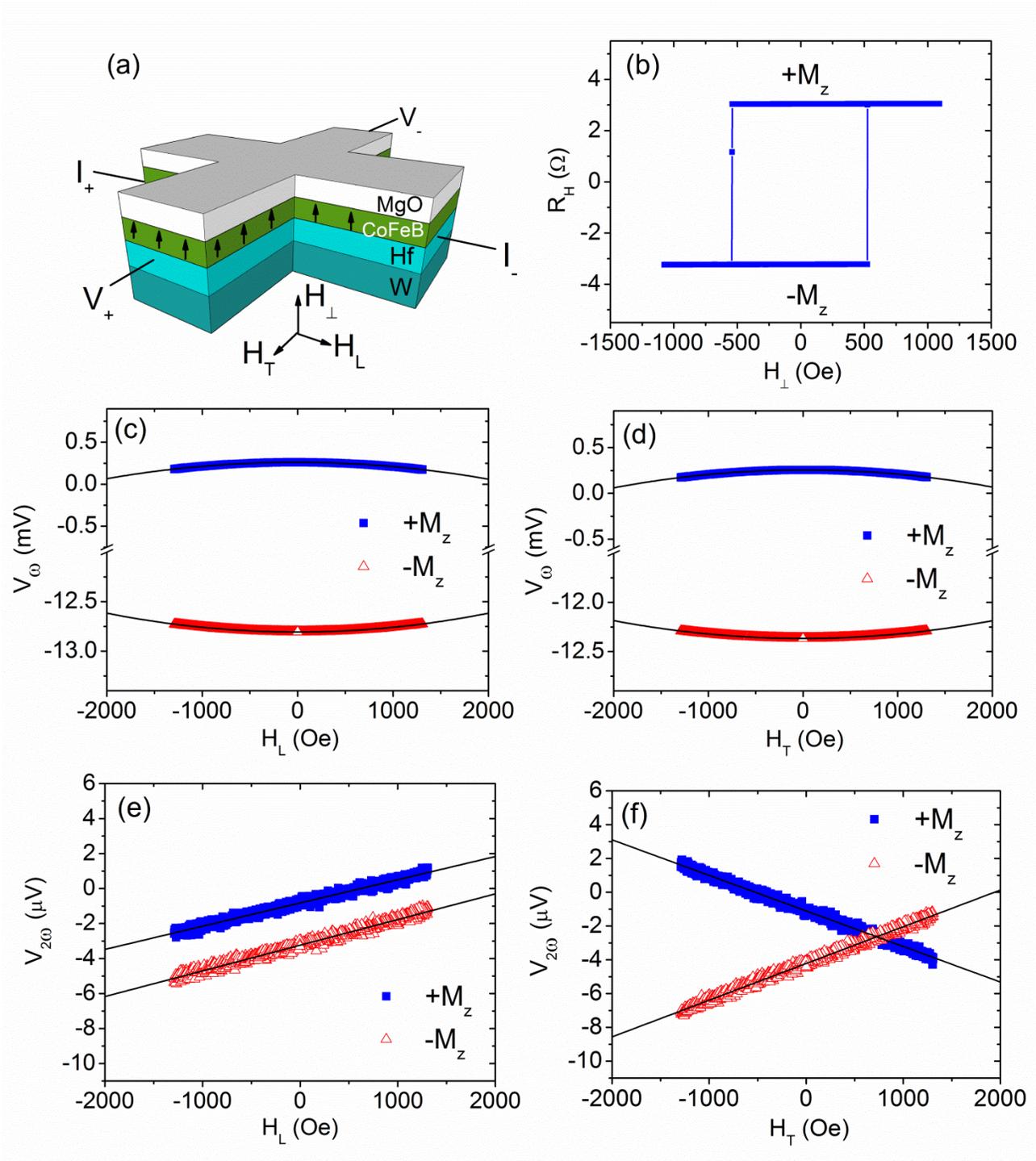

**FIG. 2**

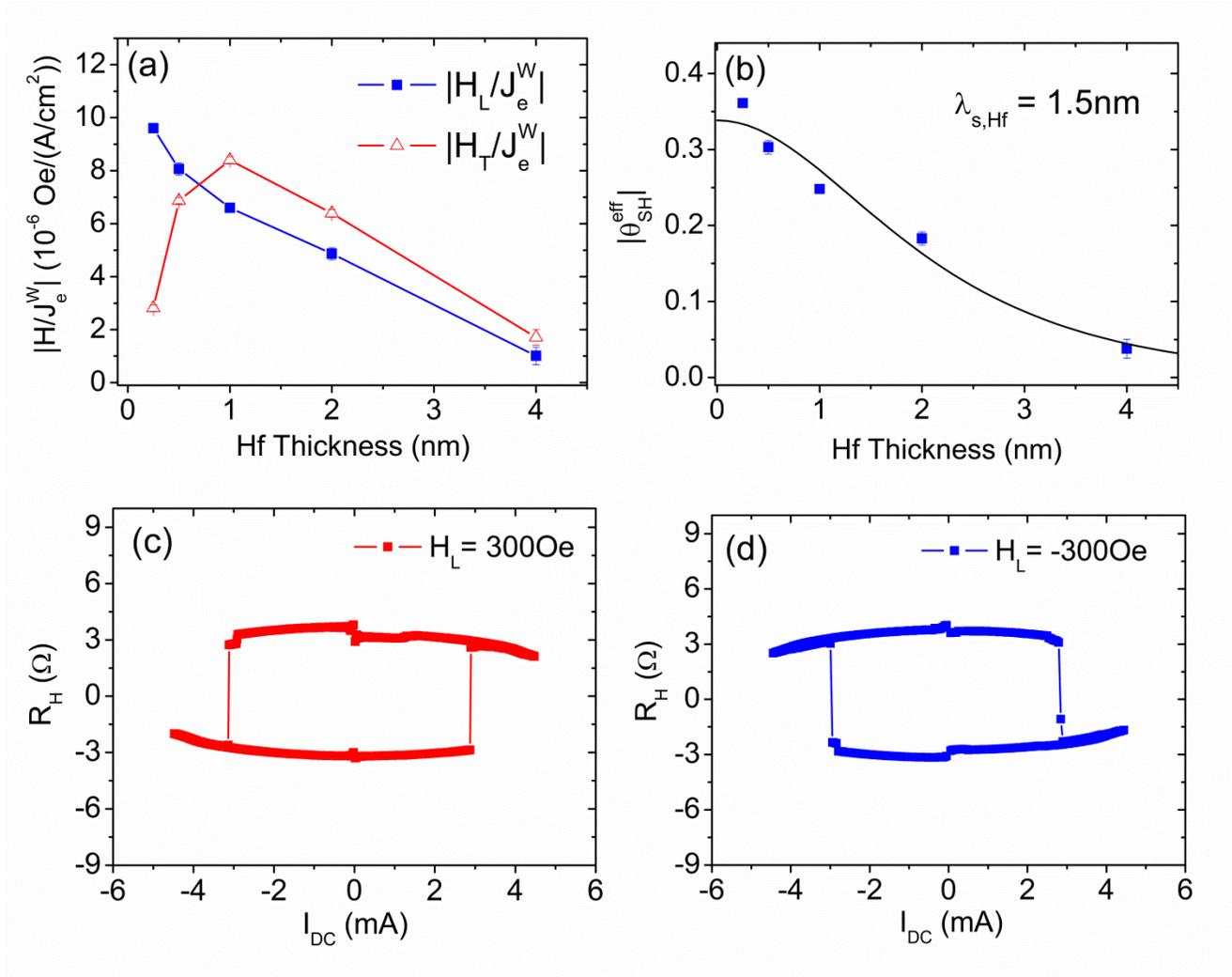

**FIG. 3**